\documentclass[12pt,preprint]{aastex}

\shorttitle{GALAXY INTERACTION AND STARBURST-SEYFERT CONNECTION}
\shortauthors{MOURI \& TANIGUCHI}

\begin{document}

\title{GALAXY INTERACTION AND STARBURST-SEYFERT CONNECTION}

\author{HIDEAKI MOURI}
\affil{Meteorological Research Institute, Nagamine 1-1, Tsukuba 305-0052, Japan; hmouri@mri-jma.go.jp}

\and

\author{YOSHIAKI TANIGUCHI}
\affil{Astronomical Institute, Graduate School of Science, Tohoku University, Aoba, Sendai 980-8578, Japan; tani@astr.tohoku.ac.jp}

\begin{abstract}

Galaxy interactions are studied in terms of the starburst-Seyfert connection. The starburst requires a high rate of gas supply. Since the efficiency for supplying the gas is high in a galaxy interaction, although the companion is not necessarily discernible, Seyfert galaxies with circumnuclear starbursts are expected to be interacting. Since the large amounts of circumnuclear gas and dust obscure the broad-line region, they are expected to be observed as Seyfert 2. The active galactic nucleus itself does not require a high rate of gas supply. Seyfert galaxies without circumnuclear starbursts are not necessarily expected to be interacting even at the highest luminosities. They are not necessarily expected to evolve from Seyfert galaxies with circumnuclear starbursts. We derive these and other theoretical expectations and confirm them with statistics on observational data of magnitude-limited samples of Seyfert galaxies. 

\end{abstract}


\notetoeditor{You may find the expression ``$10^1$'' etc. in our order-of-magnitude discussion. Please do not replace it with, e.g., ``10''.}

\section{INTRODUCTION}

When a galaxy interacts with another galaxy, its disk becomes gravitationally unstable. This instability induces a nonaxisymmetric distortion such as a bar, which drives gas of the disk into the circumnuclear region. The gas could serve as the fuel of a starburst. Owing to the interaction itself or an internal self-gravitational instability, the gas could be further driven into the nuclear region. The gas could serve as the fuel of an active galactic nucleus (AGN) if there is the central engine, a supermassive black hole (BH) (Shlosman, Begelman, \& Frank 1990; Combes 2001).

Of interest is a relation among the galaxy interaction, circumnuclear starburst, and AGN (Cid Fernandes et al. 2001; Storchi-Bergmann et al. 2001; and references therein). The relation between the latter two is called the starburst-AGN connection. We are especially interested in the relation in Seyfert galaxies, representative class of galaxies that harbor a moderate-luminosity AGN. 

While a starburst requires a higher rate of gas supply than an AGN, a galaxy interaction supplies the gas more efficiently than other mechanisms, e.g., internal self-gravitational instability of the disk. Thus a circumnuclear starburst in a Seyfert galaxy prefers an interaction as the mechanism to supply the gas (Combes 2001). We actually found that Seyfert galaxies with circumnuclear starbursts have a significant trend to be interacting while those without circumnuclear starbursts do not (Mouri \& Taniguchi 2002a; see also Storchi-Bergmann et al. 2001).

We explore the relation between galaxy interaction and the starburst-Seyfert connection in details. We derive theoretical expectations for the role of interaction, and examine whether they are consistent with statistics on observational data. This approach taken from Byrd \& Valtonen (2001) is favorable because any existing sample of Seyfert galaxies is not large or complete enough to yield statistically convincing results, and because any observational characterization of galaxy interaction has ambiguities. The observational data are described in \S2. Theories for galaxy interaction are summarized in \S3. They are compared with the data in \S4. The results of other recent observations are also discussed in \S5. We conclude with general remarks for the starburst-Seyfert connection in \S6.

Throughout this paper, we assume a Hubble constant $H_0 = 75$ km s$^{-1}$ Mpc$^{-1}$. Seyfert galaxies are of two types. Optical broad recombination lines are visible in Seyfert 1 galaxies, while they are not visible in Seyfert 2 galaxies. There are subtypes 1.0 and 1.5 in Seyfert 1 and subtypes 1.8, 1.9, and 2.0 in Seyfert 2.

\section{SAMPLES AND DATA}

Seyfert galaxies studied here are those described in our previous work (Mouri \& Taniguchi 2002a). For two magnitude-limited samples of Seyfert galaxies, far-infrared flux densities at 60 and 100 \micron\ observed by the {\it Infrared Astronomical Satellite} were used to classify the individual galaxies into those where the far-infrared continuum emission is dominated by either the AGN, circumnuclear starburst, or host galaxy. The last class is not relevant directly to the starburst-Seyfert connection, but it exists as a definite group. We outline the samples and classification scheme and then identify interacting galaxies. 

The one sample was from the Revised Shapley-Ames (RSA) Catalog of Bright Galaxies. RSA galaxies have the total $B$-band magnitudes $B_T \le 12.5$. Ho, Filippenko, \& Sargent (1997a, b) defined a sample of 52 Seyfert galaxies. They share 10\% of RSA galaxies and serve as a fair representation of the local population of Seyfert galaxies.

The other sample was from the catalog of the Center for Astrophysics (CfA) Redshift Survey. CfA galaxies have blue magnitudes $m_B \le 14.5 $ on the Zwicky-$B(0)$ system. Huchra \& Burg (1992) and Osterbrock \& Martel (1993) defined a sample of 48 Seyfert galaxies. To match the median distances for Seyfert 1 and 2 galaxies, we excluded 9 Seyfert 1 galaxies with $m_B > 14.0$. The CfA sample misses many Seyfert galaxies with weak line emission and accordingly contains a high fraction of Seyfert galaxies with strong line emission, which are rare in the RSA sample.

Our classification of Seyfert galaxies was based on the far-infrared flux between 40 and 120 \micron\ ($F_{\rm FIR}$; Fullmer \& Lonsdale 1989), the far-infrared flux densities at 60 and 100 \micron\ ($S_{60}$ and $S_{100}$ in units of janskys), and the monochromatic blue flux at 0.43 \micron\ ($F_B$; Lang 1980, p. 560):
\begin{eqnarray}
\label{eq1}
& &\mbox{AGN dominant:} \qquad \ \  
F_{\rm FIR}/F_B \le 1 \mbox{ and } S_{100}/S_{60} \le 3 \nonumber \\
& &\mbox{starburst dominant:} \quad  
F_{\rm FIR}/F_B > 1                                                \\
& &\mbox{host dominant:} \qquad \quad
S_{100}/S_{60} > 3,                                      \nonumber
\end{eqnarray}
with
\begin{equation}
\label{eq2}
F_{\rm FIR}
=
1.26 \times 10^{-11}
\left( 2.58 S_{60} + S_{100} \right) \, \mbox{ergs s}^{-1} \, \mbox{cm}^{-2}
\end{equation}
and
\begin{equation}
\label{eq3}
F_B
=
2.16 \times 10^{-5-0.4 m_B} \, \mbox{ergs s}^{-1} \, \mbox{cm}^{-2}.
\end{equation}
Starburst-dominant Seyfert galaxies have large values of $F_{\rm FIR}/F_B$, i.e., the far-infrared luminosity normalized by the blue luminosity of the galaxy. Within and around the star-forming regions, a large amount of dust grains absorbs ultraviolet photons from OB stars and reradiates their energies in the far infrared (Wilson 1988; Mouri \& Taniguchi 1992). Host-dominant Seyfert galaxies have large values of $S_{100}/S_{60}$. Since the AGN and star-forming regions are not luminous, outlying cold dust grains dominate the far-infrared emission. The thresholds were determined using the data of starburst galaxies. Mouri \& Taniguchi (2002a) tabulated the classes of the individual Seyfert galaxies together with their observational data taken from, e.g., the Third Reference Catalogue of Bright Galaxies (RC3; de Vaucouleurs et al. 1991).\footnote{
The far-infrared flux densities of some of the Seyfert galaxies were from Soifer et al. (1989) and Sanders et al. (1995). Although the data were recently revised by Sanders et al. (2003), we have ascertained that the revision does not affect the class of any of the Seyfert galaxies.} 
The number statistics are summarized in Table 1.


We search for interacting companions around each of the Seyfert galaxies. For our RSA sample, we adopt the empirical criterion of Rafanelli, Violato, \& Baruffolo (1995). A galaxy is considered as a companion if its angular distance from the Seyfert galaxy $r_{\rm sc}$ is less than 3 times the major diameter of the Seyfert galaxy $d_{\rm s}$, if the difference in the total $B$-band magnitude $\left| \delta B_T \right|$ is less than 3 mag, and if the difference in the heliocentric recession velocity $\left| \delta V_{\sun} \right|$ is less than 1000 km s$^{-1}$:
\begin{equation}
\label{eq4}
r_{\rm sc} \le 3d_{\rm s}, \ 
\left| \delta B_T \right| \le 3, \mbox{ and }
\left| \delta V_{\sun} \right| \le 1000 \, \mbox{km s}^{-1}.
\end{equation}
The recession velocity of the companion is available from RC3 or an existing redshift survey, e.g., the second CfA Redshift Survey that observed galaxies with $m_B \le 15.5$ (see Huchra, Vogeley, \& Geller 1999). For our CfA sample where galaxies tend to be fainter, we adopt a modified criterion for the companion:
\begin{equation}
\label{eq5}
r_{\rm sc} \le 3d_{\rm s}, \
m_B \le 15.5, \mbox{ and } 
\left| \delta V_{\sun} \right| \le 1000 \, \mbox{km s}^{-1}.
\end{equation}
The search is done on the NASA/IPAC Extragalactic Database. If more than one companions are found around a Seyfert galaxy, we focus on the closest companion. The RSA and CfA Seyfert galaxies for which we find a companion are listed in Tables 2 and 3, respectively.


The presence of a companion does not affect our classification of the Seyfert galaxy. Since the {\it Infrared Astronomical Satellite} had large apertures, the companion could contribute to the far-infrared flux densities observed for the Seyfert galaxy. The contribution is nevertheless small because the companion is less luminous and less active than the Seyfert galaxy (\S4.2). For example, using a high-resolution processing algorithm, Surace et al. (1993) studied images at 60 and 100 \micron\ around the Seyfert galaxy NGC 3227. Its far-infrared flux densities were found to be 30--50 times greater than those of the companion NGC 3226.

We quantify the strength of galaxy interaction using the nondimensional parameter $Q$ defined by Dahari (1984):
\begin{equation}
\label{eq6}
Q = \frac{(d_{\rm s} d_{\rm c})^{3/2}}{r_{\rm sc}^3}.
\end{equation}
Here $d_{\rm c}$ is the major diameter of the companion. The values are listed in Tables 2 and 3. All of strongly interacting companions with $Q \ga 1$ around both RSA and CfA Seyfert galaxies are expected to be captured in our search. This is not expected for weakly interacting companions with $Q \la 1$. Some of them might have been missed especially around faint CfA Seyfert galaxies. Some of objects with $Q \la 1$ captured in our search might not be proximate to the Seyfert galaxy in space but originate in the background or foreground contamination. These are unavoidable ambiguities in any observational characterization of galaxy interaction.

We also use the interaction class (IAC) determined by Dahari \& De Robertis (1988), which is available for 70\% of CfA Seyfert galaxies. The IAC scheme quantifies the strengths of galaxy interaction and large-scale distortion within the galaxy on a one-dimensional integer scale: IAC = 1 for an isolated and symmetric galaxy, and IAC = 6 for a strongly interacting and distorted galaxy. The CfA Seyfert galaxies with ${\rm IAC} = 5$ and 6 happen to be the same as those with $Q > 1$.

\section{THEORIES}

\subsection{Required Rate of Gas Supply}

Between a starburst and an AGN, the radiative efficiency is different. The luminosity of a starburst averaged over 10$^8$ yr is $\varepsilon_{\rm nuc} \varepsilon_{\rm star} \dot{M}_{\rm gas} c^2$ (Kennicutt 1998). Here $\varepsilon_{\rm nuc} \simeq 0.007$ is the fraction of mass that is released as energy in the nuclear reaction, $\varepsilon_{\rm star}$ is the fraction of stellar mass that is burned in 10$^8$ yr ($\simeq 0.05$ for Salpeter's initial mass function), and $\dot{M}_{\rm gas}$ is the star formation rate. To sustain the bolometric luminosity $L_{\rm bol}$, the rate of gas supply has to be
\begin{equation}
\label{eq7}
\frac{dM_{\rm gas}}{dt}
=
1.9 \, M_{\sun} \, {\rm yr}^{-1}
\left( \frac{\varepsilon_{\rm nuc}}  {0.007}               \right)^{-1}
\left( \frac{\varepsilon_{\rm star}} {0.05}                \right)^{-1}
\left( \frac{L_{\rm bol}}            {10^{10} \, L_{\sun}} \right).
\end{equation}
The luminosity of an AGN is $\varepsilon_{\rm BH} \dot{M}_{\rm gas} c^2$. Here $\varepsilon_{\rm BH} \simeq 0.1$ is the binding energy per unit mass of a particle in the closest stable orbit around the BH ($1-2\sqrt{2}/3$ for a Schwarzschild BH and $1-5/3\sqrt{3}$ to $1-1/\sqrt{3}$ for a Kerr BH; Shapiro \& Teukolsky 1983, pp. 346, 362, 428), and $\dot{M}_{\rm gas}$ is in this case the accretion rate. To sustain the bolometric luminosity $L_{\rm bol}$, the rate of gas supply has to be
\begin{equation}
\label{eq8}
\frac{dM_{\rm gas}}{dt}
=
0.0068 \, M_{\sun} \, {\rm yr}^{-1}
\left( \frac{\varepsilon_{\rm BH}} {0.1}       \right)^{-1}
\left( \frac{L_{\rm bol}}{10^{10} \, L_{\sun}} \right).
\end{equation}
Thus, while a high rate of gas supply is necessary for a starburst, a low rate of gas supply is sufficient for an AGN. Only when the rate of gas supply is high, the luminosity of a circumnuclear starburst dominates over that of the AGN.

\subsection{Fuel Source and Fueling Efficiency}

The rate of gas supply is determined by two parameters: (1) the amount of available gas within the galaxy and (2) the efficiency for transporting the gas into the circumnuclear or nuclear region. The first parameter is related to the morphological type of the galaxy. A later-type galaxy has a larger amount of gas (Young \& Scoville 1991; Roberts \& Haynes 1994). Although this is not the case if the morphological type is later than about Scd, such a small-bulge or bulgeless galaxy is rarely observed as Seyfert because the central supermassive BH is rarely present (Ho et al. 1997b; \S3.5). The second parameter is related to the mechanism that generates a distortion and thereby induces gas inflow. Various mechanisms with various efficiencies are available (Shlosman et al. 1990; Combes 2001; see also Wada 2004). Galaxy interactions are the most efficient.

The amount of gas and the efficiency for transporting the gas put serious constraints on the luminosity of a starburst, which requires a high rate of gas supply. The constraints are not serious to the luminosity of an AGN (Combes 2001), which does not require a high rate of gas supply. For a moderate-luminosity AGN in a Seyfert galaxy, a small amount of gas as in an elliptical galaxy and a low transport efficiency as in a weak internal self-gravitational instability are sufficient. It is unnecessary to gather gas from the entire galaxy. Such an inefficient mechanism is not evident in the large-scale appearance of the galaxy.\footnote{
Xilouris \& Papadakis (2002) reported that RSA Seyfert galaxies always exhibit a distortion in their central $10^{-1}$--$10^0$ kpc regions while RSA absorption-line galaxies do not always exhibit such a distortion. That is probably the largest-scale common signature of gas supply in Seyfert galaxies.} 

To the rate of gas supply, the amount of gas is more important than the efficiency for transporting the gas. For example, if the galaxy is of a late type and has a large amount of gas, even an inefficient mechanism supplies the gas at a high rate. The morphological type also affects consequences of a galaxy interaction. Since a massive bulge stabilizes the disk against the tidal perturbation and thereby suppresses gas inflow (Mihos \& Hernquist 1994; Hernquist \& Mihos 1995), the transport of gas is more efficient in a later-type galaxy where the bulge is less massive.

Early-type galaxies tend to interact frequently because they tend to reside in dense environments, e.g., galaxy clusters (Roberts \& Haynes 1994). However, among Seyfert galaxies, the fraction of interacting objects is not accordingly dependent on the morphological type. Galaxies in a dense environment tend to have high relative velocities. In a high-velocity interaction, the distortion and hence the gas inflow tend to be negligible because the tidal perturbation is impulsive and is not resonant with the internal dynamics of the galaxy. The companion quickly escapes from the vicinity of the galaxy (see also \S3.4).

\subsection{Merger and Transient Encounter}

A galaxy interaction is either a merger or a transient encounter. In a merger, the companion merges with the galaxy due to dynamical friction. In a transient encounter, the companion passes away from the galaxy. A major merger with a comparable-mass galaxy induces a strong tidal perturbation and a high rate of gas supply. Even if the bulge is massive and stabilizes the disk against the perturbation (\S3.2), this stabilization is eventually overwhelmed by the merger (Mihos \& Hernquist 1994). A minor merger with a small satellite galaxy induces a weaker tidal perturbation and a lower rate of gas supply. The rate of gas supply is also lower in a transient encounter because the companion is relatively distant from the Seyfert galaxy. We do not consider transient encounters with a small galaxy, where the rate of gas supply is negligibly low.

Observational data are not sufficient to determine the entire character of the interaction, but the parameter $Q$ in equation (\ref{eq6}) offers some insights. Since the masses $M_{\rm s}$ and $M_{\rm c}$ of the Seyfert galaxy and the companion are approximately related to their diameters $D_{\rm s}$ and $D_{\rm c}$ as $M_{\rm s} \propto D^{3/2}_{\rm s}$ and $M_{\rm c} \propto D^{3/2}_{\rm c}$ (Rubin et al. 1982), the parameter $Q$ is regarded as the strength of the tidal perturbation normalized by the self-gravity of the Seyfert galaxy (Byrd \& Valtonen 2001):
\begin{equation}
\label{eq9}
Q
=
\frac{(d_{\rm s} d_{\rm c})^{3/2}}{r_{\rm sc}^3}
\simeq
\frac{(D_{\rm s} D_{\rm c})^{3/2}}{R_{\rm sc}^3}
\simeq
\frac{M_{\rm c} D_{\rm s} / R_{\rm sc}^3}{M_{\rm s} / D_{\rm s}^2}.
\end{equation}
Here $R_{\rm sc}$ is the distance in space between the Seyfert galaxy and the companion. In addition, a companion with $Q \ga 1$ is ready to merge with the Seyfert galaxy (see Byrd \& Valtonen 2001). We separate strong interactions with $Q > 1$ from weak interactions with $Q \le 1$. A strong interaction roughly corresponds to a major merger, while a weak interaction roughly corresponds to a minor merger or a transient encounter.

\subsection{Relevant Timescales}

The timescales that are relevant to gas supply are related to dynamical timescales. They are obtained by dividing representative scales by representative velocities. Although the gas supply has temporal variations and feedbacks, e.g., removal of gas due to radiation from the starburst or AGN, these timescales are of fundamental importance.

The duration $\tau_{\rm dur}$ of gas supply in a transient encounter is comparable to the dynamical timescale $\tau_{\rm dyn}$ for the entire galaxy (Byrd \& Valtonen 2001):
\begin{equation}
\label{eq10}
\tau_{\rm dur}
\simeq
\tau_{\rm dyn}
=
6.5 \times 10^7 \, \mbox{yr}
\left( \frac{D_{\rm s}}      {10 \,  \mbox{kpc}}        \right)
\left( \frac{ V_{\rm orb,s} }{150 \, \mbox{km s}^{-1} } \right)^{-1}.
\end{equation}
Here $V_{\rm orb,s}$ is the typical orbital velocity for gas within the Seyfert galaxy. In a major merger, the duration is comparable to the merger timescale that is approximately the orbital period $2 \pi R_{\rm sc}/V_{\rm orb,sc}$ times the mass ratio $M_{\rm heavy}/M_{\rm light}$ (Carico et al. 1990):
\begin{equation}
\label{eq11}
\tau_{\rm dur}
\simeq
\tau_{\rm dyn}
=
6.1 \times 10^8  \, \mbox{yr}
\left( \frac{R_{\rm sc}}     {30 \, \mbox{kpc}}        \right)
\left( \frac{V_{\rm orb,sc}} {300 \, \mbox{km s}^{-1}} \right)^{-1}
\left( \frac{M_{\rm heavy}}  {M_{\rm light}}           \right).
\end{equation}
Here $V_{\rm orb,sc}$ is the orbital velocity between the Seyfert galaxy and the companion, $M_{\rm heavy}$ is the mass of the more massive object of the two, and $M_{\rm light}$ is the mass of the less massive object. In a minor merger, the duration is shorter because gas supply does not occur until the companion sinks into the central few kiloparsecs of the galaxy (Hernquist \& Mihos 1995).

The delay $\tau_{\rm del}$ for the gas to arrive at the nuclear region after the onset of transient encounter or major merger is comparable to the dynamical timescale for the entire galaxy (Byrd \& Valtonen 2001):
\begin{equation}
\label{eq12}
\tau_{\rm del}
\simeq
\tau_{\rm dyn}
=
6.5 \times 10^7 \, \mbox{yr}
\left( \frac{D_{\rm s}}      {10 \, \mbox{kpc}}         \right)
\left( \frac{ V_{\rm orb,s} }{150 \, \mbox{km s}^{-1} } \right)^{-1}.
\end{equation}
The delay is longer in a minor merger where gas supply occurs only at the final stage. Since a transient encounter has $\tau_{\rm del} \simeq \tau_{\rm dur}$ and a minor merger has $\tau_{\rm del} \ga \tau_{\rm dur}$, there is a large number of inactive interacting galaxies. The delay also makes it difficult to find the companion. In a transient encounter, the companion could have escaped from the vicinity of the Seyfert galaxy (Byrd \& Valtonen 2001; Combes 2001). This is because the change of distance during the delay $\delta R_{\rm sc}$ could be comparable to our search radius for the companion (\S2, eqs. [\ref{eq4}] and [\ref{eq5}]):
\begin{equation}
\label{eq13}
\delta R_{\rm sc}
\simeq 
V_{\rm rel,sc} \tau_{\rm del}
\simeq
2 D_{\rm s}
\left( \frac{ V_{\rm rel,sc}}{300 \, \mbox{km s}^{-1}} \right)
\left( \frac{ V_{\rm orb,s} }{150 \, \mbox{km s}^{-1} } \right)^{-1}.
\end{equation}
Here $V_{\rm rel,sc}$ is the relative velocity between the Seyfert galaxy and the companion. In a minor merger, the companion could have been absorbed into the Seyfert galaxy (Hernquist \& Mihos 1995; Taniguchi \& Wada 1996).\footnote{
De Robertis, Yee, \& Hayhoe (1998) and Taniguchi (1999) argued that a minor merger is the predominant mechanism for gas supply in apparently noninteracting Seyfert galaxies, but our discussion in \S3.2 and \S3.4 implies that other various mechanisms such as a transient encounter and internal self-gravitational instabilities are equally important.} 
Nevertheless, in a major merger, the delay is unimportant because the merger timescale is longer by an order of magnitude.

\subsection{Mass of Central Black Hole}

Whether a circumnuclear starburst outshines the AGN depends not only on the rate of gas supply but also on the mass of the central BH. This is because the luminosity of an AGN is constrained by the BH mass $M_{\rm BH}$ through the Eddington limit $L_{\rm Edd}$ (Shapiro \& Teukolsky 1983, p. 395):
\begin{equation}
\label{eq14}
L_{\rm bol} 
\le
L_{\rm Edd}
=
3.3 \times 10^{11} \, L_{\sun} \, 
\left( \frac{M_{\rm BH}}{10^7 \, M_{\sun}} \right).
\end{equation}
Seyfert galaxies have $L_{\rm bol}/L_{\rm Edd} \simeq 10^{-2}$--$10^{-1}$. Quasars have $L_{\rm bol}/L_{\rm Edd} \ga 10^{-1}$. The BH mass ranges from $10^6$ to $10^{10} \, M_{\sun}$ and scales with the mass of the galaxy bulge (Kormendy \& Richstone 1995; Kormendy \& Gebhardt 2001). In a later-type galaxy, the bulge mass and hence the BH mass are smaller. The Eddington limit is accordingly lower.

The origin of the central BH is controversial and beyond the scope of our study (see \S6 for one of the possible origins). We assume the preexistence of the central BH. Its mass increases via accretion during the AGN activity and via mergers with other supermassive BHs. The companion in a major or minor merger could harbor such a BH.

\section{THEORIES VERSUS OBSERVATIONS}

Observational data of our Seyfert galaxies that are to be compared with our theoretical expectations are summarized in Figures 1--3 and Tables 1--3. The figures show numbers of AGN-, starburst-, and host-dominant objects as a function of the morphological type. The numbers of all the objects are also shown. In Figures 1 and 2, filled areas denote strongly interacting galaxies with $Q > 1$, grayed areas denote weakly interacting galaxies with $Q \le 1$, and open areas denote galaxies for which we do not find a companion. The total fraction of interacting galaxies is independent of the morphological type (\S3.2), especially in the RSA sample where the identifications of companions are more reliable (\S2). In Figure 3, filled areas denote strongly interacting or distorted galaxies with IAC = 5--6, grayed areas denote weakly interacting or distorted galaxies with IAC = 2--4, and open areas denote isolated and symmetric galaxies with IAC = 1.


\subsection{Predominance of Circumnuclear Starburst}

Starburst-dominant Seyfert galaxies are expected to have later morphological types than AGN-dominant Seyfert galaxies. This dependence on morphological type is expected to be more significant than the dependence on interaction and distortion properties. A starburst requires a high rate of gas supply (\S3.1). The rate is determined primarily by the amount of available gas, which is larger in a later-type galaxy (\S3.2). On the other hand, the possible maximum luminosity of the AGN is determined by the mass of the central BH, which is smaller in a later-type galaxy (\S3.5).

Unless the morphological type is very early and hence the amount of gas is very small, strongly interacting Seyfert galaxies are expected to be dominated by a circumnuclear starburst. The strong interaction efficiently transports gas into the circumnuclear region (\S3.3). In addition, the strong interaction such as a major merger has a long duration of gas supply so that its delay after the onset of interaction is statistically unimportant (\S3.4).

Weakly interacting Seyfert galaxies are not necessarily expected to be dominated by a circumnuclear starburst. The rate of gas supply might not be sufficient to induce a luminous starburst. This is especially the case in an early-type galaxy where gas is deficient (\S3.2). The AGN might not be fueled by the observed interaction. This is the case where the interaction has not yet transported gas into the circumnuclear or nuclear region (\S3.4).
 
Seyfert galaxies that appear to be noninteracting are expected to be found not only among those dominated by the AGN or host galaxy but also among those dominated by a circumnuclear starburst. The gas that is responsible for the starburst could have been supplied by a mechanism other than an interaction, especially in a late-type galaxy where gas is abundant (\S3.2). The companion in a weak interaction is difficult to discern (\S3.4). We nevertheless expect the presence of a large-scale distortion that corresponds to a high rate of gas supply in any starburst-dominant Seyfert galaxy. 

Figures 1--3 show that starburst-dominant objects are of later types than AGN-dominant objects (Storchi-Bergmann et al. 2001; Mouri \& Taniguchi 2002a). Strongly interacting or distorted galaxies with $Q > 1$ or IAC = 5--6 exist only among starburst-dominant objects ({\it filled areas}). In Figures 1 and 2, many of starburst-dominant objects are weakly interacting with $Q \le 1$ or apparently noninteracting ({\it grayed and open areas}). Nevertheless, in Figure 3, almost all of starburst-dominant objects exhibit a large-scale distortion, ${\rm IAC} = 2$--6 ({\it filled and grayed areas}). Only some of AGN- and host-dominant objects exhibit no large-scale distortion, ${\rm IAC} = 1$ ({\it open areas}).

\subsection{Companion}

The companion is expected to have a smaller size and a lower luminosity than the Seyfert galaxy. The central engine of an AGN, a supermassive BH, prefers a massive galaxy with a massive bulge (\S3.5). Since more massive galaxies are less numerous, if the relative masses of interaction pairs are random, the companion tends to be less massive than the Seyfert galaxy. A less massive galaxy has a smaller size and a lower luminosity.

The companion is expected to exhibit or have exhibited an activity that is less prominent than that of the Seyfert galaxy. Since the parameter $Q$ for the companion is the same as that for the Seyfert galaxy, the companion suffers from the same strength of tidal perturbation if normalized by the self-gravity (Byrd \& Valtonen 2001; \S3.3, eq. [\ref{eq9}]). Since the companion is less massive than the Seyfert galaxy, the amount of available gas and the mass of the central BH are smaller (\S3.5). Since the companion is smaller than the Seyfert galaxy, the delay and duration of gas supply are shorter (\S3.4).

RSA and CfA Seyfert galaxies are luminous. If the Galactic extinction $A_B = 0.12 \csc b$ at the latitude $b$ is assumed, the median absolute magnitudes on the Zwicky-$B(0)$ system for our RSA and CfA samples are $-20.1$ and $-20.2$ mag, respectively. The mean absolute magnitudes are $-19.8$ and $-20.3$ mag, respectively (see Mouri \& Taniguchi 2002a). A typical galaxy has $-19.7$ mag (Schechter 1976). Tables 2 and 3 show that almost every companion has a smaller diameter and a lower brightness than the Seyfert galaxy (Storchi-Bergmann et al. 2001). Several of the companions are known to harbor a LINER, i.e., low-luminosity AGN, or a starburst (NGC 3034, NGC 3073, NGC 3226, NGC 4217, and NGC 5195; Ho et al. 1997a), but none of them is known to harbor a moderate- or high-luminosity AGN.

\subsection{Seyfert Type}

According to the so-called unified model (Antonucci 1993), Seyfert 1 and 2 galaxies are identical but are observed in different orientations. Around  the central BH and the broad-line region, there is a dusty torus. The torus obscures the broad-line region along our line of sight to a Seyfert 2 galaxy. This model holds for galaxies where observational properties are determined by the AGN alone, i.e., AGN-dominant Seyfert galaxies.

The situation is different in strongly interacting Seyfert galaxies. We expect that these galaxies tend to be observed as Seyfert 2, regardless of the orientation of the torus. The strong interaction transports large amounts of gas and dust into the circumnuclear region. They obscure the broad-line region along most lines of sight.

Strongly interacting objects with $Q > 1$ in our CfA sample tend to be observed as Seyfert 2: 4 in 5 objects (Table 3). This fraction of Seyfert 2 galaxies is higher than that among AGN-dominant objects: 10 in 18 objects (Table 1). The result for our RSA sample is not inconsistent. The fraction of Seyfert 2 galaxies is 2 in 3 strongly interacting objects while it is 9 in 15 AGN-dominant objects (Tables 1 and 2). Table 1 also shows that the fraction of Seyfert 2 galaxies is high among starburst-dominant Seyfert galaxies: 12 in 14 RSA objects and 13 in 17 CfA objects. There are large amounts of obscuring gas and dust associated with the circumnuclear starburst (Mouri \& Taniguchi 2002a; see also Heckman et al. 1989).

Pogge \& Martini (2002) studied high-resolution images of CfA Seyfert galaxies obtained by the {\it Hubble Space Telescope}. In strongly interacting galaxies with $Q > 1$, the circumnuclear or nuclear region was found to be dusty. A strong interaction is surely capable of transporting large amounts of gas and dust to obscure the broad-line region.

\subsection{Luminosity}

Strongly interacting objects are expected to share a higher fraction in a sample of more luminous Seyfert galaxies. Since a strong interaction efficiently transports gas (\S3.3), the star formation rate is high. The accretion rate onto the AGN tends to be also high so far as its luminosity is well below the Eddington limit (\S3.5). Our CfA sample contains a higher fraction of strongly interacting objects with $Q > 1$ than our RSA sample: 5 in 39 CfA objects and 3 in 52 RSA objects (Tables 1--3). CfA Seyfert galaxies tend to exhibit stronger line emission than RSA Seyfert galaxies (\S2).

The correspondence with a strong interaction is expected to be significant in the far-infrared luminosity, which is sensitive to the circumnuclear starburst (\S2). An example is luminous infrared galaxies with the far-infrared luminosities $L_{\rm FIR} \ga 10^{11} \, L_{\sun}$. Sanders \& Mirabel (1996) showed the presence of a high fraction of mergers and close pairs. The AGNs where the broad lines are not visible are more common than those where the broad lines are visible.

The highest luminosity, $L_{\rm bol} \gg 10^{12}\, L_{\sun}$, is nevertheless expected for objects dominated by an AGN because its radiative efficiency is $10^2$ times higher than that of a starburst (\S3.1, eqs. [{\ref{eq7}] and [\ref{eq8}]). Even among the luminous infrared galaxies, with an increase of the far-infrared luminosity, the fraction of broad-line objects and hence the fraction of AGN-dominant objects increase (Sanders \& Mirabel 1996; see also \S4.3). The AGN luminosity is constrained by the mass of the central BH, which is large in a galaxy with a massive bulge (\S3.5). Dunlop et al. (2003) found that host galaxies of luminous quasars are very massive but otherwise normal ellipticals. They are not necessarily undergoing a strong interaction. A weak interaction or an internal mechanism is sufficient for supplying gas to the quasar because a very massive elliptical has a relatively large amount of available gas.

\subsection{Frequency of Gas Supply}

The probability for a Seyfert activity to be induced by a major merger in a galaxy with the central BH is $\sim 1$\% per $10^9$ yr. This is because 10\% of RSA galaxies are Seyfert galaxies (Ho et al. 1997b) and 6\% of them are strongly interacting with $Q > 1$ (Tables 1 and 2). The fraction of strongly-interacting Seyfert galaxies among all galaxies is thereby of the order 1\%. This fraction is of the same order as the probability times the duration, which is $10^9$ yr for a major merger (\S3.4, eq. [\ref{eq11}]). The presence of galaxies without the central BHs is unimportant here because they are not the majority (see Ho et al. 1997b).

The probability for a Seyfert activity to be induced by an episode of gas supply other than a major merger in a galaxy with the central BH is $\ga 100$\% per $10^9$ yr, judging from the fraction of Seyfert galaxies in the RSA sample, 10\%. Such an episode is responsible for most of the Seyfert galaxies and has a duration that does not exceed the dynamical timescale $10^8$ yr for the entire galaxy (\S3.4). Since the present age of a galaxy is $10^{10}$ yr, it follows that a Seyfert activity is repetitive and has occurred several times on average in a galaxy with the central BH (Mouri \& Taniguchi 2002a).

\subsection{Typical Evolution}

When a late-type galaxy undergoes a transient encounter, gas is supplied from the disk to the circumnuclear region at a high rate and induces a luminous starburst. Then the gas is transported into the nuclear region. The starburst galaxy is expected to become a starburst-dominant Seyfert galaxy if there is the central BH. The delay $\tau_{\rm del}$ of the onset of AGN after the onset of starburst is comparable to the dynamical timescale for the circumnuclear region with the size $D'_{\rm s}$:
\begin{equation}
\label{eq15}
\tau_{\rm del}
\simeq
\tau_{\rm dyn}
=
6.5 \times 10^6 \, \mbox{yr}
\left( \frac{D'_{\rm s}}    {1 \, \mbox{kpc}}          \right)
\left( \frac{V_{\rm orb,s} }{150 \, \mbox{km s}^{-1} } \right)^{-1}.
\end{equation}
The duration of gas supply is 10$^8$ yr (\S3.4, eq. [\ref{eq10}]). Thereafter the starburst and AGN become less luminous. The object is expected to become a host-dominant Seyfert galaxy,\footnote{
Host-dominant objects have not been used in our discussion because they are not relevant directly to the starburst-Seyfert connection. Figures 1--3 show that the fraction of host-dominant objects is higher in our less luminous RSA sample than it is in our more luminous CfA sample. The morphological types of host-dominant objects cover those of AGN- and starburst-dominant objects. Host-dominant objects are not strongly interacting or distorted. These are because, in a host-dominant object, the rate of gas supply is too low or the mass of the central BH is too small for the starburst or AGN to outshine the host galaxy. Table 1 shows that host-dominant objects consist mostly of Seyfert 2 galaxies: 17 in 18 RSA objects. In addition to the obscuration by the dusty torus or circumnuclear gas and dust, the obscuration by the interstellar medium of the host galaxy and the dilution by the stellar absorption features hamper the detection of their weak broad lines. This is especially the case in edge-on objects. There is also a possibility that the broad-line region is intrinsically absent in some of such weak AGNs. See also Mouri \& Taniguchi (2002a).} 
before it becomes an inactive galaxy. The object does not become an AGN-dominant Seyfert galaxy because the central BH is not massive enough (\S3.5). Throughout the evolution, the object remains a late-type galaxy. More time is required for the morphological type to change markedly. A similar evolution is expected when a late-type galaxy undergoes a minor merger.

When an early-type galaxy undergoes a transient encounter or a minor merger, the rate of gas supply is not high. The object is expected to become an AGN- or host-dominant Seyfert galaxy if there is the central BH. The AGN-dominant Seyfert galaxy is expected to become a host-dominant Seyfert galaxy.

When a galaxy undergoes a major merger, the rate of gas supply is high if the amount of available gas is not very small. The object is expected to become a starburst galaxy $\rightarrow$ a starburst-dominant Seyfert galaxy if there is the central BH. Since the duration is as long as 10$^9$ yr (\S3.4, eq. [\ref{eq11}]), all the gas within the galaxy is transported into the circumnuclear and nuclear regions and is consumed by the starburst and AGN. The accretion onto the BH increases its mass and hence the possible maximum luminosity of the AGN (\S3.5). Probably via an AGN-dominant stage, the object is expected to become a host-dominant Seyfert galaxy. If the initial amount of available gas is very small, the evolution starts from the AGN- or host-dominant stage. The object eventually becomes a gas-deficient elliptical. An example is the merger remnant NGC 7252, which has star clusters with ages $\la 10^9$ yr (Whitmore et al. 1993; see also Sanders \& Mirabel 1996).

These expectations are consistent with statistical properties of AGN-, starburst-, and host-dominant Seyfert galaxies in Figures 1--3. In addition, the far-infrared colors $S_{100}/S_{60}$ of RSA and CfA starburst-dominant Seyfert galaxies are close to those expected for starbursts with ages up to several times 10$^8$ yr (Mouri \& Taniguchi 2002a). Starbursts with similar ages were found in other observations of Seyfert galaxies, e.g., optical colors and absorption features (Cid Fernandes et al. 2001), near-infrared colors (Hunt et al. 1997), and equivalent width of the Br$\gamma$ line (Oliva et al. 1995). Very young starbursts with ages $\la 10^6$ yr are rare in Seyfert galaxies (Glass \& Moorwood 1985; Mouri \& Taniguchi 1992) because of the delay of the onset of AGN after the onset of starburst (eq. [\ref{eq15}]).\footnote{
For the presence of a very young starburst, the one possibility is that the starburst region is close to the nucleus and hence the delay of the onset of AGN was short. The other possibility is that numerous OB stars were recently formed owing to, e.g., new supply of a large amount of gas.}

\section{INTERPRETATION OF RECENT OBSERVATIONS}

De Robertis et al. (1998) studied the two-point correlation function around individual CfA Seyfert galaxies. The correlation function $\xi (R_{\rm sc}) = (R_{\rm sc}/R_0)^{-1.77} = \xi_0 R_{\rm sc}^{-1.77}$ gives the excess probability for finding a companion at the projected distance $R_{\rm sc}$ from the galaxy. It was found that Seyfert 2 galaxies tend to have larger amplitudes $\xi_0$ than Seyfert 1 galaxies. The interaction with the companion could transport large amounts of gas and dust into the circumnuclear region and thereby obscure the broad-line region (\S4.3). The gas could also induce a starburst. Seyfert 2 galaxies with large amplitudes $\xi_0$ are actually dominated by a circumnuclear starburst (Mouri \& Taniguchi 2002a).

Dultzin-Hacyan et al. (1999) studied Seyfert galaxies selected from an ultraviolet survey. Compared with Seyfert 1 galaxies, Seyfert 2 galaxies were found to exhibit an excess of nearby large companions. This excess is again due to interacting starburst-dominant Seyfert 2 galaxies. They are preferred over AGN-dominant Seyfert 2 galaxies in an ultraviolet survey (Schmitt et al. 2001). In the former objects, the ultraviolet emission arises from the starburst. In the latter objects, there is no strong source of ultraviolet emission.

Schmitt et al. (2001) studied companions around Seyfert galaxies with redshifts $z \le 0.031$ selected on the basis of their warm 25--60 \micron\ colors. No systematic difference was found between Seyfert 1 and 2 galaxies. The warm 25--60 \micron\ color implies that the sample consists preferentially of AGN-dominant Seyfert galaxies (Wilson 1988; Mouri \& Taniguchi 1992). Objects where the broad-line region is obscured by circumnuclear gas and dust, e.g., interacting starburst-dominant Seyfert 2 galaxies, are rare in the sample.

Chatzichristou (2002) studied Seyfert galaxies that were selected with the same criterion as in Schmitt et al. (2001) for warm 25--60 \micron\ colors but from a wider range of redshift $z = 0.01$--0.08. Strongly interacting objects were found to share a higher fraction among Seyfert 2 galaxies than they were among Seyfert 1 galaxies. Compared with the sample of Schmitt et al. (2001), there is a high fraction of far-infrared luminous objects up to $L_{\rm FIR} \simeq 10^{12} \, L_{\sun}$. Most of them are interacting starburst-dominant Seyfert 2 galaxies (\S4.4). Their warm 25--60 \micron\ colors are attributable to young starbursts where dust grains are heated to high temperatures by hot O stars (Mouri \& Taniguchi 1992).

Schmitt (2001) studied companions around RSA galaxies. No systematic difference was found between Seyfert galaxies and relatively inactive galaxies, e.g., LINERs and absorption-line galaxies. A similar result for CfA galaxies was obtained by De Robertis et al. (1998). Seyfert galaxies as a whole do not exhibit a significant trend to be interacting, especially in a sample of low- and moderate-luminosity galaxies. The interaction is not necessary for gas supply if its required rate is not very high (\S3.2). The companion is not necessarily discernible (\S3.4). Also, there are inactive interacting galaxies, e.g., objects where the interaction has not yet transported gas into the nuclear region (\S3.4).

Kauffmann et al. (2003) studied Seyfert 2 galaxies selected from the Sloan Digital Sky Survey. Close companions or significant tidal debris were found for 30\% of objects with the [\ion{O}{3}] 0.5007 $\micron$ luminosities greater than about $10^{8}\ L_{\sun}$. This fraction of strongly interacting objects is higher than those for $Q > 1$ among RSA and CfA Seyfert 2 galaxies: 2 in 43 RSA objects and 4 in 26 CfA objects (Tables 1--3). Strongly interacting objects share a higher fraction in a sample of more luminous Seyfert galaxies (\S4.4). The median [\ion{O}{3}] luminosity for RSA Seyfert galaxies is $10^6\ L_{\sun}$ (see Ho et al. 1997b).

Therefore, the fraction of interacting objects in a sample is sensitive to its selection criteria for the spectral energy distribution and luminosity. Since the sensitivity is mainly via the fraction of starburst-dominant objects, it is important to separate these objects as in our present work (see also Storchi-Bergmann et al. 2001; Mouri \& Taniguchi 2002a).

\section{CONCLUDING SUMMARY}

Thus far we have explored the relation between galaxy interaction and the starburst-Seyfert connection. We have derived theoretical expectations for the role of interaction and confirmed them with statistics on observational data of RSA and CfA Seyfert galaxies. They had been classified into those where the far-infrared continuum emission is dominated by either the AGN, circumnuclear starburst, or host galaxy (Mouri \& Taniguchi 2002a). We have also discussed the results of other recent observations. Here we conclude with general remarks for the starburst-Seyfert connection. The main concepts are summarized in Figure 4.


To sustain the luminosity, a starburst requires a higher rate of gas supply than an AGN (\S3.1). The rate of gas supply is determined by the amount of available gas within the galaxy and the efficiency for transporting the gas into the circumnuclear or nuclear region (\S3.2).

The amount of gas is larger in a later-type galaxy (\S3.2). Since the central BH is less massive, the possible maximum luminosity of the AGN is lower (\S3.5). Starburst-dominant Seyfert galaxies are of later types than AGN-dominant Seyfert galaxies (\S4.1).

The most efficient mechanism for gas supply is strong interactions such as a major merger. The major merger also has the longest duration of gas supply, $10^9$ yr (\S3.4). Strongly interacting Seyfert galaxies are found preferentially among starburst-dominant objects (\S4.1).

Weak interactions such as a minor merger and a transient encounter are less efficient. The duration of gas supply is shorter and does not exceed its delay after the onset of interaction. The companion is difficult to discern (\S3.4). There is no significant correspondence between a weak interaction and a starburst-dominant Seyfert galaxy (\S4.1).
 
The least efficient mechanism, i.e., internal self-gravitational instability, is still sufficient for a moderate-luminosity AGN (\S3.2). This mechanism is difficult to be identified because of its intrinsic insignificance and its decay during the transport of gas into the nuclear region. Some of AGN- and host-dominant Seyfert galaxies do not exhibit any large-scale distortion (\S4.1; see also Combes 2001).
 
Compared with inactive galaxies, Seyfert galaxies as a whole do not exhibit a significant trend to be interacting (\S5). The interaction is not necessary for gas supply. The companion is not necessarily discernible. Also, there are inactive interacting galaxies (\S3.2 and \S3.4).

The companion suffers from the same strength of tidal perturbation as the Seyfert galaxy if the perturbation is normalized by the self-gravity. However, since the Seyfert galaxy tends to be massive, the companion tends to be less massive and accordingly less active (\S4.2). 

The broad-line region of a Seyfert galaxy could be obscured by gas and dust that have been transported into the circumnuclear region by a galaxy interaction or an internal mechanism. The gas could also induce a starburst. Thus an object that is strongly interacting or dominated by a circumnuclear starburst tends to be observed as Seyfert 2 (\S4.3; Mouri \& Taniguchi 2002a). 

Since AGN-, starburst-, and host-dominant Seyfert galaxies have different spectral energy distributions and luminosities, the fraction of interacting starburst-dominant Seyfert 2 galaxies in a sample is sensitive to its selection criteria. This fact is responsible for dissimilar results among recent studies to compare the fraction of interacting objects between Seyfert 1 and 2 galaxies (\S5).

The importance of strong interaction is greater among more luminous objects, especially in the far-infrared continuum. However, the most luminous objects are not necessarily undergoing a strong interaction and are dominated by an AGN because it is $10^2$ times more efficient in generating radiation than a starburst (\S4.4, see also \S3.1 and \S5).

The duration of gas supply and hence the duration of the Seyfert activity with or without a circumnuclear starburst do not exceed the dynamical timescale 10$^8$ yr for the entire galaxy if the gas is supplied by a mechanism other than a major merger (\S3.4). These durations are too short for the morphological type of the galaxy to change markedly (\S4.6). 

The activity with a duration $\la 10^8$ yr is responsible for most of Seyfert galaxies, which are not undergoing a major merger. Such an activity is repetitive and has occurred several times on average in a galaxy with the central BH. This is because Seyfert galaxies share 10\% of all galaxies. The present age of a galaxy is 10$^{10}$ yr (\S4.5; Mouri \& Taniguchi 2002a).  

The typical evolution of a Seyfert galaxy is modelled as follows (\S4.6; Mouri \& Taniguchi 2002a). We regard starburst and Seyfert activities as ingredients of the evolution of a galaxy from a late type to an early type, and propose a model that is in accordance with the current paradigm for evolution of the galaxy bulge (Carollo 2004). Suppose a late-type galaxy that harbors the central BH. When gas is supplied at a high rate by a minor merger, a transient encounter, or an internal mechanism, the galaxy evolves to starburst $\rightarrow$ starburst-dominant Seyfert $\rightarrow$ host-dominant Seyfert $\rightarrow$ inactive. This sequence is repeated several times. The repetition of starburst and Seyfert activities consumes gas in the galaxy and increases the mass of the BH. While the possible maximum luminosity of the starburst decreases, that of the AGN increases. The galaxy becomes to have an early-type morphology and becomes to evolve to AGN-dominant Seyfert $\rightarrow$ host-dominant Seyfert $\rightarrow$ inactive. A sequence starts from the host-dominant Seyfert stage if the rate of gas supply is very low. The exception is a major merger, which singly produces the full evolution.

The above model ignores effects of the megaparsec-scale environment (\S3.2). They are not completely understood and would be an important subject in the future researches.

Some of the existing models assume that an AGN-dominant Seyfert galaxy evolves from a starburst-dominant Seyfert galaxy during a single episode of gas supply. Such an evolution could be expected for a major merger (\S4.6), but it is not typical. A modest episode of gas supply that is not sufficient for a starburst but is sufficient for a moderate-luminosity AGN occurs more frequently (\S4.5).

Finally, we comment on galaxies without the central supermassive BHs, e.g., very late-type spirals. The object becomes a starburst galaxy when gas is supplied at a high rate to the circumnuclear or nuclear region. This is the same situation as for circumnuclear starbursts in Seyfert galaxies. Luminous starburst galaxies actually tend to be interacting (Kennicutt 1998). Within the individual star clusters of the starburst region, massive stars and stellar-mass BHs merge with each other and form intermediate-mass BHs ($10^0\, M_{\sun} \ll M_{\rm BH} \ll 10^6\, M_{\sun}$; Taniguchi et al. 2000). The star clusters sink into the galaxy center due to dynamical friction and deposit the intermediate-mass BHs. They merge with each other and form a supermassive BH (Ebisuzaki et al. 2001). Under favorable conditions, an intermediate-mass BH is formed in $\la 10^7$ yr, and a star cluster sinks into the galaxy center in 10$^8$--10$^9$ yr (Mouri \& Taniguchi 2002b, 2003). The starburst galaxy becomes a starburst-dominant Seyfert galaxy if the duration of gas supply is long as in the case of major merger. Even if this is not the case, the galaxy becomes a Seyfert galaxy when gas is supplied again in the future.

\acknowledgments
The NASA/IPAC Extragalactic Database is operated by the Jet Propulsion Laboratory, California Institute of Technology, under contract with the National Aeronautics and Space Administration. We are grateful for financial support from the Japanese Ministry of Education, Science, and Culture under grants 10044052 and 10304013 and for useful comments of the referee.


\clearpage

\begin{deluxetable}{lccccc}

\tabletypesize{\scriptsize}
\tablenum{1}
\tablecolumns{6}
\tablewidth{0pc}
\tablecaption{Galaxy Numbers in RSA and CfA Samples}

\tablehead{
\colhead{}              &
\multicolumn{2}{c}{RSA} &
\colhead{}              &
\multicolumn{2}{c}{CfA} \\ \cline{2-3} \cline{5-6}
\colhead{Class} &
\colhead{S1}   &
\colhead{S2}   &
\colhead{}     &
\colhead{S1}   &
\colhead{S2} }

\startdata
AGN dominant       & 6 &  9 & & 8  & 10 \\
Starburst dominant & 2 & 12 & & 4  & 13 \\
Host dominant      & 1 & 17 & & 1  &  1 \\
Unclassified       & 0 &  5 & & 0  &  2 \\
All                & 9 & 43 & & 13 & 26 
\enddata
\tablecomments{S1 = Seyfert 1 (Seyfert 1.0 and 1.5). S2 = Seyfert 2 (Seyfert 1.8, 1.9, and 2.0). The data of the individual galaxies are given in Mouri \& Taniguchi (2002a).}
\end{deluxetable}

\begin{deluxetable}{lccrcrrclrrcrcrrc}

\tabletypesize{\scriptsize}
\setlength{\tabcolsep}{0.02in}
\tablenum{2}
\tablecolumns{17}
\tablewidth{0pc}
\tablecaption{POSSIBLE COMPANIONS AROUND RSA SEYFERT GALAXIES}

\tablehead{
\multicolumn{7}{c}{Seyfert Galaxy\tablenotemark{a}} &
\colhead{}                                          &
\multicolumn{9}{c}{Possible Companion} \\ \cline{1-7} \cline{9-17}
\colhead{}                 &
\multicolumn{2}{c}{Class}  &
\colhead{}                 &
\colhead{}                 &
\colhead{}                 &
\colhead{}                 &
\colhead{}                 &
\colhead{}                 &
\colhead{}                 &
\colhead{}                 &
\colhead{}                 &
\colhead{}                 &
\colhead{}                 &
\colhead{}                 &
\colhead{}                 &
\colhead{}                 \\ \cline{2-3}
\colhead{Name}             &
\colhead{Opt.}             &
\colhead{FIR}              & 
\colhead{$T$}              &
\colhead{$\log d_{\rm s}$} &
\colhead{$B_T$}            &
\colhead{$V_{\sun}$}       &
\colhead{}                 &
\colhead{Name}             &
\colhead{$\alpha$(1950)}   &
\colhead{$\delta$(1950)}   &
\colhead{$\log r_{\rm sc}$}&
\colhead{$T$}              &
\colhead{$\log  d_{\rm c}$}&
\colhead{$B_T$}            &
\colhead{$V_{\sun}$}       &
\colhead{$Q$}             \\
\colhead{(1)} &
\colhead{(2)} &
\colhead{(3)} &
\colhead{(4)} &
\colhead{(5)} &
\colhead{(6)} &
\colhead{(7)} &
\colhead{}    &
\colhead{(8)} &
\colhead{(9)} &
\colhead{(10)}&
\colhead{(11)}&
\colhead{(12)}&
\colhead{(13)}&
\colhead{(14)}&
\colhead{(15)}&
\colhead{(16)} }

\startdata
NGC 777\tablenotemark{b}  & 2.0 & \nodata  & $-5.0$ & 1.39 & 12.49    &  4985 & & NGC 778\tablenotemark{c}   & 01 57 25.6 & $+31$ 04 17 & 1.85 & $-2.0$       & 1.03 & \nodata    & 5433 & 0.01  \\
NGC 1275\tablenotemark{d} & 1.5 & S        &  99.0  & 1.34 & 12.64    &  5260 & & NGC 1278                   & 03 16 35.7 & $+41$ 23 00 & 1.53 & $-5.0$       & 1.19 & 13.57      & 6047 & 0.16  \\
NGC 1358\tablenotemark{d} & 2.0 & A        &  0.0   & 1.41 & 13.04    &  4021 & & NGC 1355                   & 03 30 54.6 & $-05$ 09 58 & 1.83 & $-1.9$       & 1.16 & 14.25      & 3915 & 0.02  \\
NGC 3031                  & 1.5 & H        &  2.0   & 2.43 &  7.89    & $-49$ & & NGC 3034                   & 09 51 45.3 & $+69$ 55 11 & 2.57 & 90.0         & 2.05 &  9.30      &  300 & 0.10  \\
NGC 3079                  & 2.0 & S        &  7.0   & 1.90 & 11.54    &  1101 & & NGC 3073                   & 09 57 28.8 & $+55$ 51 39 & 2.23 & $-2.5$       & 1.11 & 14.07      & 1176 & 0.01  \\
NGC 3227                  & 1.5 & S        &  1.0   & 1.73 & 11.1\phn &  1145 & & NGC 3226                   & 10 20 43.6 & $+20$ 09 07 & 1.37 & $-5.0$       & 1.50 & 12.3\phn   & 1322 & 5.46  \\
NGC 4168\tablenotemark{b} & 1.9 & \nodata  & $-5.0$ & 1.44 & 12.11    &  2284 & & NGC 4165                   & 12 09 38.9 & $+13$ 31 29 & 1.43 & 1.0          & 1.12 & 14.38      & 1523 & 0.37  \\
NGC 4169\tablenotemark{d} & 2.0 & H        & $-2.0$ & 1.26 & 13.15    &  3783 & & NGC 4174                   & 12 09 55.1 & $+29$ 25 38 & 1.44 & $-2$\phn\phd & 0.92 & 14.31      & 3980 & 0.09  \\
NGC 4258                  & 1.9 & H        &  4.0   & 2.27 &  9.10    &   480 & & NGC 4217                   & 12 13 21.7 & $+47$ 22 12 & 2.69 & 3.0          & 1.72 & 12.04      & 1032 & 0.01  \\
NGC 4477                  & 2.0 & A        & $-2.0$ & 1.58 & 11.38    &  1348 & & NGC 4479                   & 12 27 46.8 & $+13$ 51 15 & 1.73 & $-2.0$       & 1.19 & 13.40      &  822 & 0.09  \\
NGC 4725                  & 2.0 & H        &  2.0   & 2.03 & 10.11    &  1180 & & NGC 4747                   & 12 49 18.7 & $+26$ 02 45 & 2.41 & 6.0          & 1.54 & 12.96      & 1219 & 0.01  \\
NGC 5033                  & 1.9 & S        &  5.0   & 2.03 & 10.75    &   861 & & UGC 8303                   & 13 10 59.2 & $+36$ 28 44 & 2.36 & 10.0         & 1.35 & 13.48      &  944 & 0.01  \\
NGC 5194                  & 2.0 & S        &  4.0   & 2.05 &  8.96    &   463 & & NGC 5195                   & 13 27 52.5 & $+47$ 31 48 & 1.68 & 90.0         & 1.76 & 10.45      &  570 & 4.86  \\
NGC 5395                  & 2.0 & S        &  3.0   & 1.46 & 12.1\phn &  3505 & & NGC 5394                   & 13 56 25.2 & $+37$ 41 51 & 1.32 & 3.0          & 1.24 & 13.7\phn   & 3427 & 1.22  \\
\enddata

\tablecomments{Col. (1): Object name. Col. (2): Classification based on the optical line emission, i.e., Seyfert type. Col. (3): Classification based on the far-infrared continuum emission (A = AGN dominant; S = starburst dominant; H = host dominant). Col. (4): Morphological type index as defined in RC3. Col. (5): Decimal logarithm of major isophotal diameter at the $B$-band surface brightness 25 mag arcsec$^{-2}$ in units of 0.1 arcminutes taken from RC3. Col. (6): Total $B$-band magnitude taken from RC3. Col. (7): Heliocentric recession velocity in units of km s$^{-1}$. Col. (8): Object name. Cols. (9) and (10): Position in equatorial coordinates taken from RC3 (1950 equinox). Units of right ascension are hours, minutes, and seconds. Units of declination are degrees, arcminutes, and arcseconds. Col. (11): Decimal logarithm of distance from the Seyfert galaxy in units of 0.1 arcminutes. Col. (12) Morphological type index as defined in RC3 taken from RC3 or Huchra et al. (1990). Col. (13): Decimal logarithm of major isophotal diameter at the $B$-band surface brightness 25 mag arcsec$^{-2}$ in units of 0.1 arcminutes taken from RC3. Col. (14): Total $B$-band magnitude taken from RC3. Col. (15): Heliocentric recession velocity in units of km s$^{-1}$ taken from RC3, Huchra et al. (1999, 1995, 1983), or da Costa et al. (1998). Col. (16): Interaction strength defined as $Q = (d_{\rm s}d_{\rm c})^{3/2}/r_{\rm sc}^3$.}

\tablenotetext{a}{Unless otherwise noted here, references for the data are given in Mouri \& Taniguchi (2002a).}
\tablenotetext{b}{Since the far-infrared emission was not detected, the far-infrared classification was not possible.}
\tablenotetext{c}{The total $B$-band magnitude is unavailable, but the Zwicky-$B(0)$ magnitude of NGC 778 is close to that of NGC 777 (14.2 and 13.09 mag, respectively; Huchra et al. 1999).}
\tablenotetext{d}{This galaxy is out of the formal RSA range $B_T \le 12.5$, but it was included in the sample of Ho et al. (1997a, b). The reason is a special or historical interest or an error in the magnitude given in the RSA catalog.}

\end{deluxetable}
\notetoeditor{Please do not change the order of the above references, which indicates the decreasing order of preference.}
\notetoeditor{Please do not replace, e.g., "11.1" with "11.10" because there are "12.49", "9.10", ... in the same column. This is not a typological error. The accuracies are different.}

\begin{deluxetable}{lccrclcclcccrcllc}

\tabletypesize{\scriptsize}
\setlength{\tabcolsep}{0.02in}
\tablenum{3}
\tablecolumns{17}
\tablewidth{0pc}
\tablecaption{POSSIBLE COMPANIONS AROUND CFA SEYFERT GALAXIES}

\tablehead{
\multicolumn{7}{c}{Seyfert Galaxy\tablenotemark{a}} &
\colhead{}                                          &
\multicolumn{9}{c}{Possible Companion} \\ \cline{1-7} \cline{9-17}
\colhead{}                 &
\multicolumn{2}{c}{Class}  &
\colhead{}                 &
\colhead{}                 &
\colhead{}                 &
\colhead{}                 &
\colhead{}                 &
\colhead{}                 &
\colhead{}                 &
\colhead{}                 &
\colhead{}                 &
\colhead{}                 &
\colhead{}                 &
\colhead{}                 &
\colhead{}                 &
\colhead{}                \\ \cline{2-3}
\colhead{Name}             &
\colhead{Opt.}             &
\colhead{FIR}              & 
\colhead{$T$}              &
\colhead{$\log d_{\rm s}$} &
\colhead{$m_B$}            &
\colhead{$V_{\sun}$}       &
\colhead{}                 &
\colhead{Name}             &
\colhead{$\alpha$(1950)}   &
\colhead{$\delta$(1950)}   &
\colhead{$\log r_{\rm sc}$}&
\colhead{$T$}              &
\colhead{$\log d_{\rm c}$} &
\colhead{$m_B$}            &
\colhead{$V_{\sun}$}       &
\colhead{$Q$}             \\
\colhead{(1)} &
\colhead{(2)} &
\colhead{(3)} &
\colhead{(4)} &
\colhead{(5)} &
\colhead{(6)} &
\colhead{(7)} &
\colhead{}    &
\colhead{(8)} &
\colhead{(9)} &
\colhead{(10)}&
\colhead{(11)}&
\colhead{(12)}&
\colhead{(13)}&
\colhead{(14)}&
\colhead{(15)}&
\colhead{(16)} }

\startdata
NGC 1144                  & 2.0 & S       & 10\phd\phn & 1.04 & 13.6  &    8647 & & NGC 1143        & 02 52 36.2 & $-00$ 22 47 & 0.82    & $-4.3$  & 0.96    & 14.2    &    8459 & 3.41 \\
NGC 3227                  & 1.5 & S       &  1.0       & 1.73 & 11.75 &    1145 & & NGC 3226        & 10 20 43.6 & $+20$ 09 07 & 1.37    & $-5.0$  & 1.50    & 12.77   &    1322 & 5.46 \\
NGC 3786\tablenotemark{b} & 1.8 & \nodata &  1.0       & 1.34 & 13.5  &    2722 & & NGC 3788        & 11 37 06.4 & $+32$ 12 35 & 1.16    & 2.0     & 1.33    & 13.2    &    2486 & 3.45 \\
NGC 5033                  & 1.9 & S       &  5.0       & 2.03 & 10.85 & \phn861 & & UGC 8303        & 13 10 59.2 & $+36$ 28 44 & 2.36    & 10.0    & 1.35    & 14.7    & \phn944 & 0.01 \\
NGC 5256\tablenotemark{c} & 2.0 & S       & 99.0       & 1.08 & 14.1  &    8353 & & \phm{00}\nodata & \nodata    & \nodata     & \nodata & \nodata & \nodata & \nodata & \nodata & $\gg 1$ \\
NGC 5929                  & 2.0 & S       &  2.0       & 0.98 & 14.0  &    2514 & & NGC 5930        & 15 24 20.6 & $+41$ 51 05 & 0.83    & 3.0     & 1.22    & 13.6    &    2672 & 6.29 \\
NGC 7469                  & 1.0 & S       &  1.0       & 1.17 & 13.0  &    4846 & & IC 5283         & 23 00 47.1 & $+08$ 37 26 & 1.12    & 6.0     & 0.90    & 15.2    &    4894 & 0.57 \\
NGC 7674                  & 2.0 & S       &  4.0       & 1.05 & 13.6  &    8662 & & NGC 7675        & 23 25 33.8 & $+08$ 29 34 & 1.38    & $-3.0$  & 0.82    & 14.8    &    8616 & 0.05 \\
\enddata

\tablecomments{Cols. (1)--(5), (7)--(13), (15), and (16): See notes to Table 2. Cols. (6) and (14): Magnitude on the Zwicky-$B(0)$ system. The values for the companions were taken from Huchra et al. (1999, 1995, 1990, 1983).}

\tablenotetext{a}{Unless otherwise noted here or in Table 2, references for the data are given in Mouri \& Taniguchi (2002a).}
\tablenotetext{b}{Since the far-infrared emission was not measured, the far-infrared classification was not possible.}
\tablenotetext{c}{This is a galaxy with double nuclei.}

\end{deluxetable}
\notetoeditor{Please do not change the order of the above references, which indicates decreasing order of preference.}
\notetoeditor{Please do not replace, e.g., "13.6" with "13.60" because there are "11.75", "10.85", ... in the same column. This is not a typological error. The accuracies are different.}

\clearpage

\clearpage
\begin{figure}
\epsscale{0.7}
\plotone{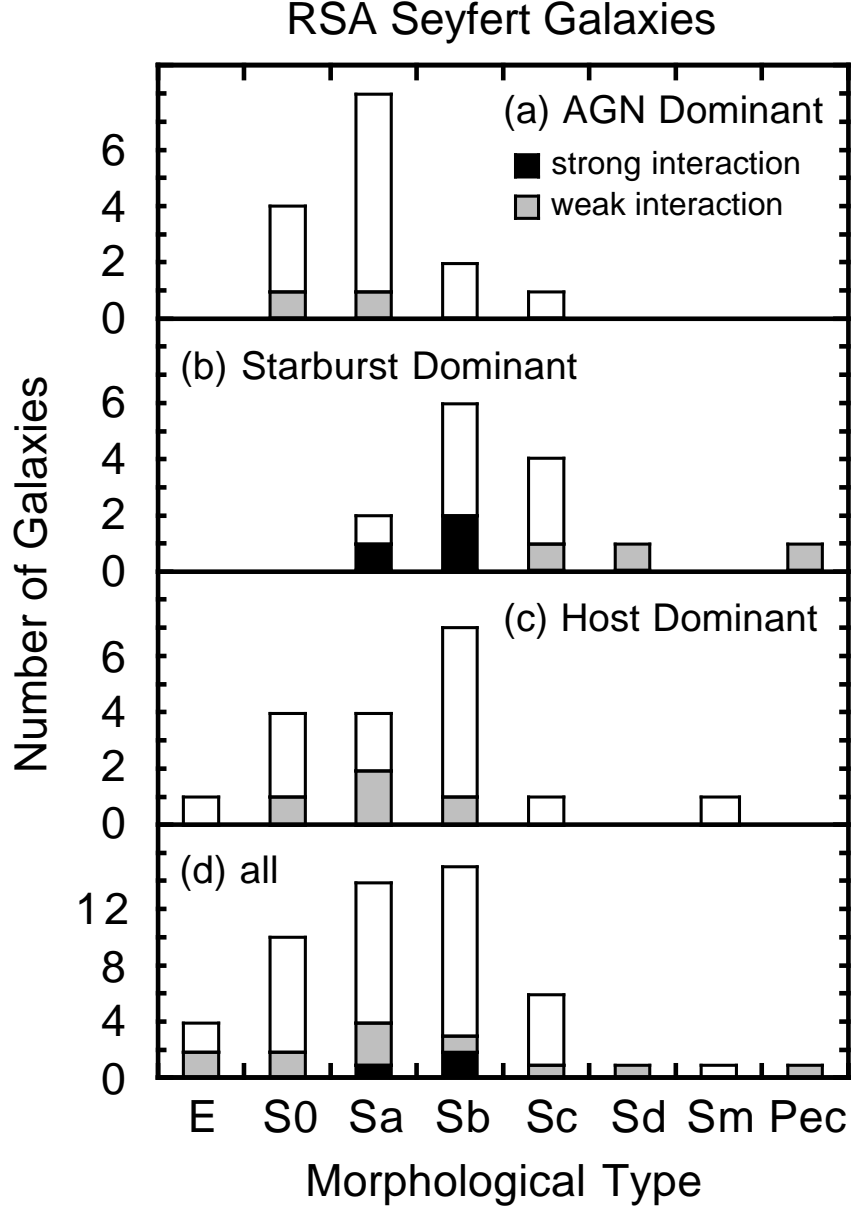}
\caption[Fig1]{Number distribution of the morphological type for RSA Seyfert galaxies. ($a$) AGN-dominant objects. ($b$) Starburst-dominant objects. ($c$) Host-dominant objects. ($d$) All the objects that include unclassified objects. {\it Filled areas}: Strongly interacting objects with $Q > 1$. {\it Grayed areas}: Weakly interacting objects with $Q \le 1$. {\it Open areas}: Noninteracting objects. The bins along the abscissa have the following meanings: ``E'' = E, ``S0'' = S0, ``Sa'' = S0/a--Sab, ``Sb'' = Sb--Sbc, ``Sc'' = Sc--Scd, ``Sd'' = Sd--Sdm, ``Sm'' = Sm--Im, and ``Pec'' = Pec. We plot all the objects listed in Mouri \& Taniguchi (2002a; see also our Tables 1 and 2).}
\end{figure}
\notetoeditor{"Tables 1 and 2" in the above caption are those of the present manuscript.}

\clearpage
\begin{figure}
\epsscale{0.7}
\plotone{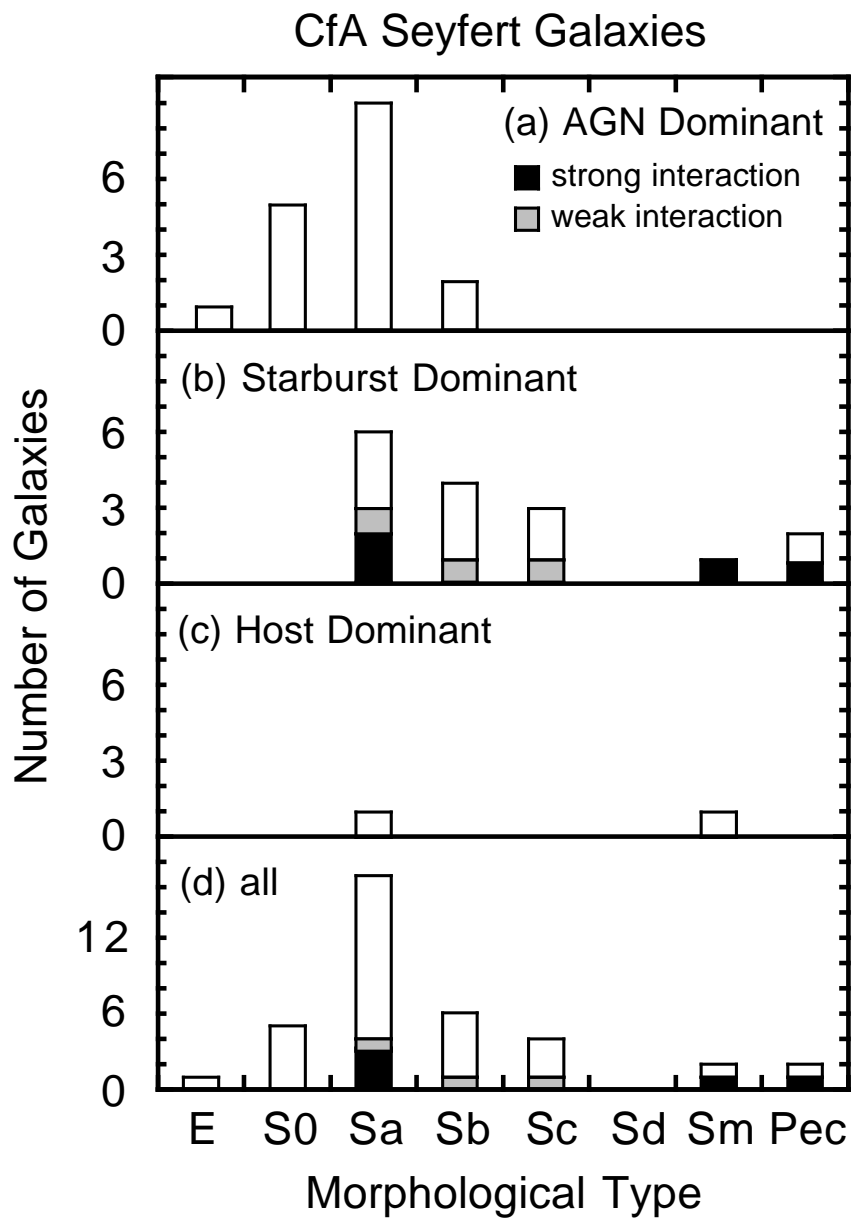}
\caption[Fig2]{Number distribution of the morphological type for CfA Seyfert galaxies. ($a$) AGN-dominant objects. ($b$) Starburst-dominant objects. ($c$) Host-dominant objects. ($d$) All the objects that include unclassified objects. {\it Filled areas}: Strongly interacting objects with $Q > 1$. {\it Grayed areas}: Weakly interacting objects with $Q \le 1$. {\it Open areas}: Noninteracting objects. The abscissa is the same as in Fig. 1. We plot all the objects listed in Mouri \& Taniguchi (2002a; see also our Tables 1 and 3), except for 1335$+$39 and Mrk 841 where the morphological type is unavailable.}
\end{figure}
\notetoeditor{"Tables 1 and 3" in the above caption are those of the present manuscript.}

\clearpage
\begin{figure}
\epsscale{0.7}
\plotone{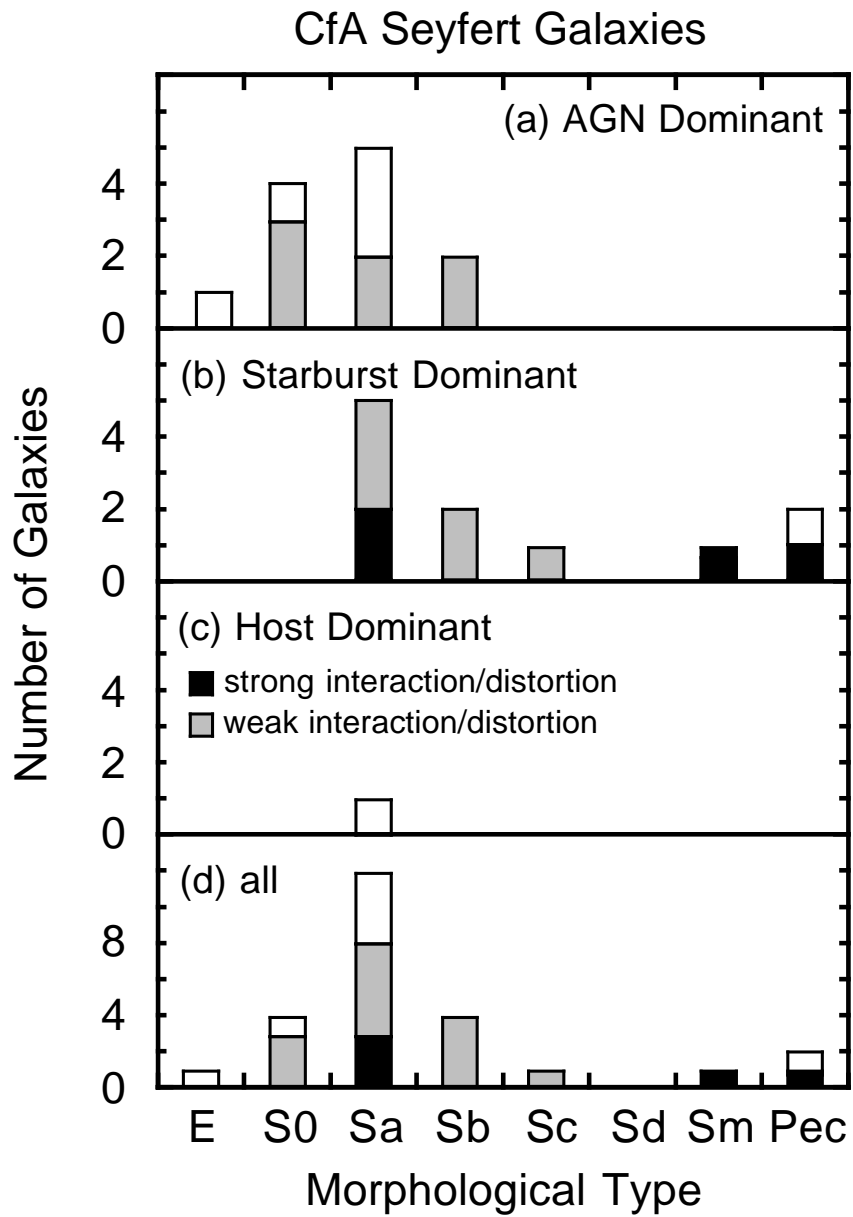}
\caption[Fig3]{Number distribution of the morphological type for CfA Seyfert galaxies where the interaction class (IAC) determined by Dahari \& De Robertis (1988) is available. ($a$) AGN-dominant objects. ($b$) Starburst-dominant objects. ($c$) Host-dominant objects. ($d$) All the objects that include unclassified objects. {\it Filled areas}: Strongly interacting or distorted objects with IAC = 5--6. {\it Grayed areas}: Weakly interacting or distorted objects with IAC = 2--4. {\it Open areas}: Isolated and symmetric objects with IAC = 1. The abscissa is the same as in Fig. 1. We plot Mrk 334, Mrk 335, Mrk 993, Mrk 573, 0152$+$06 (= UM 146), NGC 863, NGC 1068, NGC 1144, NGC 3227, NGC 3516, NGC 3786, NGC 4051, NGC 4151, NGC 4235, NGC 4253, NGC 5033, NGC 5256, NGC 5283, NGC 5273, NGC 5548, Mrk 471, NGC 5695, NGC 5929, NGC 7469, and NGC 7674.}
\end{figure}

\clearpage
\begin{figure}
\epsscale{0.8}
\plotone{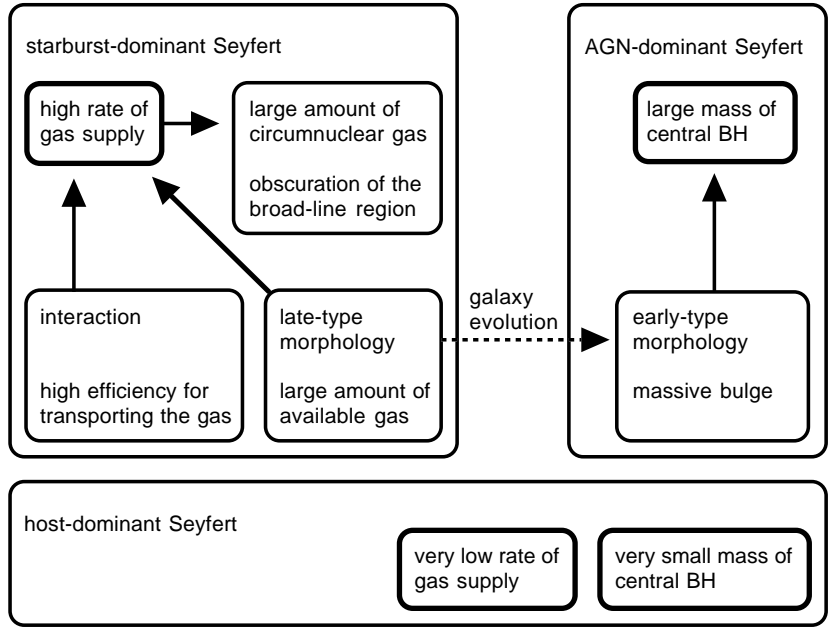}
\caption[Fig4]{Main concepts of our model for AGN-, starburst-, and host-dominant Seyfert galaxies. In an AGN-dominant Seyfert galaxy, the mass of the central BH is large. In a starburst-dominant Seyfert galaxy, the rate of gas supply is high owing to a large amount of available gas or a high efficiency for transporting the gas. In a host-dominant Seyfert galaxy, the mass of the central BH is very small or the rate of gas supply is very low.}
\end{figure}

\end{document}